\documentclass[aps,prl,twocolumn,showpacs,preprintnumbers,amsmath,amssymb,
epsf,superscriptaddress,groupedaddress, tightenlines]{revtex4}
\usepackage{graphicx}
\usepackage{amsmath}
\usepackage{bm}
\begin{document}

\preprint{\vbox{ \hbox{ANL-HEP-PR-03-102}
                 \hbox{KAIST-TH 2003/11}
                 \hbox{KIAS-P03085}
         }}
\title{%\mbox{}\\[4pt]
Chiral perturbation theory for pentaquark baryons \\
and its applications}


\author{P. Ko}
\email[]{pko@muon.kaist.ac.kr}
%\thanks{}
\affiliation{Department of Physics, KAIST, Daejon 305-701, Korea}

\author{Jungil Lee}
\email[]{jungil@hep.anl.gov}
%\thanks{}
%\alt
%\affiliation{Department of Physics, KAIST, Daejon 305-701, Korea}
\affiliation{Korea Institute for Advanced Study, Seoul 130-722, Korea}
\affiliation{High Energy Physics Division, Argonne National Laboratory, 
Argonne, Illinois 60439, U.S.A.}

\author{Taekoon Lee}
\email[]{tlee@phya.snu.ac.kr}
%\thanks{}
\affiliation{School of Physics, Seoul National University,
Seoul 151-742, Korea}

\author{Jae-hyeon Park}
\email[]{jhpark@muon.kaist.ac.kr}
%\thanks{}
\affiliation{Department of Physics, KAIST, Daejon 305-701, Korea}


\date{\today}

\begin{abstract}
We construct a chiral lagrangian for pentaquark baryons assuming that the
recently found $\Theta^+ (1540)$ state belongs to an 
antidecuplet of $\textrm{SU}(3)$ flavor symmetry with 
$J^P = \frac{1}{2}^{\pm}$. 
We derive the Gell-Mann-Okubo formulae for the antidecuplet baryon masses,
and a possible mixing between the antidecuplet and the pentaquark octet.  
Then we calculate the cross sections for 
$\pi^- p \to K^- \Theta^+$ and $\gamma n \rightarrow K^- \Theta^+$ 
using our chiral lagrangian. The resulting amplitudes respect 
the underlying chiral symmetry of QCD correctly.  
We also describe how to include the light vector mesons in the chiral 
lagrangian.
\end{abstract}

%12.39.Fe Chiral Lagrangians
%14.20.-c Baryons (including antiparticles)
%13.75.-n Hadron-induced low- and intermediate-energy reactions 
%         and scattering (energy(less-than-or-equal-to)10 GeV) 
%        (for higher energies, see 13.85.-t)
%13.60.-r photon-hadron interaction

%\keywords{}

\maketitle

%\tighten
%\section{Introduction}
%\subsection{introduction}
%\subsubsection{}

1. \textit{Introduction} --- Recently,  
five independent experiments reported observations of 
a new baryonic state $\Theta^+ (1540)$ with a very narrow width $< 5$ MeV
\cite{Nakano:2003qx,Barmin:2003vv,Stepanyan:2003qr,Kubarovsky:2003nn,
Barth:ja,Asratyan:2003cb}, which is 
likely to be a pentaquark state ($uudd\bar{s}$) \cite{Diakonov:1997mm}.  
Arguments based on quark models suggest that this state 
is a member of $\textrm{SU}(3)$ antidecuplet with spin $J=\frac{1}{2}$ or 
$\frac{3}{2}$. 
The hadro/photo production cross section would depend on the spin $J$ and 
parity  $P$ of the $\Theta^+$, and it is important to have reliable 
predictions for these cross sections. 
The most proper way to address these issues will be chiral perturbation 
theory. 

In this paper, we construct a chiral lagrangian for pentaquark baryons
assuming they are $\textrm{SU}(3)$ antidecuplet with $J=\frac{1}{2}$ 
and $P=+1$ or $-1$. 
(The case for $J=\frac{3}{2}$ can be discussed in a similar manner, 
except that antidecuplets are described by Rarita--Schwinger fields.) 
Then we calculate the mass spectra of antidecuplets, their possible 
mixings with pentaquark octets,  the decay rates of antidecuplets, and 
cross sections for 
$\pi^- p \rightarrow K^-\Theta^+$ and $\gamma n \rightarrow K^- \Theta^+$.  
Finally we describe how to include light vector mesons in our framework, 
and how the low energy theorem is recovered in the soft pion limit.

2. \textit{Chiral lagrangian for a pentaquark baryon decuplet} ---
Let us denote the Goldstone boson field by pion octet $\pi$, 
baryon octet including
nucleons by $B$, and antidecuplet including $\Theta^+$ by ${\cal P}$.
Under chiral $\textrm{SU}(3)_L\times\textrm{SU}(3)_R$ \cite{Manohar:1996cq},  
the Goldstone boson field 
$\Sigma \equiv \exp ( 2 i \pi / f ) $,
where $f\approx 93$~MeV is the pion decay constant,
transforms as
\[
\Sigma (x) \rightarrow L \Sigma(x) R^{\dagger}.
\]
It is convenient to define another field $\xi (x)$ by $\Sigma ( x ) 
\equiv \xi^2 (x) $, which transforms as
\[
\xi (x) \rightarrow L \xi (x) U^{\dagger} (x) = U(x) \xi (x) R^{\dagger} .
\] 
The $3\times 3$ matrix field $U(x)$ depends on Goldstone fields $\pi (x)$ 
as well as the $\textrm{SU}(3)$ transformation matrices $L$ and $R$. 
It is convenient to define two vector fields with following properties
under chiral transformations:
\begin{eqnarray}
V_{\mu} = 
\mbox{$\frac{1}{2}$}
\,( \xi^\dagger \partial_\mu \xi + \xi \partial_\mu  \xi^\dagger ), 
&& V_\mu \to  U V_\mu U^{\dagger}  + U \partial_\mu U^{\dagger},
\nonumber\\
A_{\mu} =
\mbox{$\frac{i}{2}$}
\,( \xi^\dagger \partial_\mu \xi - \xi \partial_\mu \xi^\dagger),
&&A_\mu \to  U A_\mu U^{\dagger}.
\end{eqnarray}
Note that $V_\mu$ transforms like a gauge field. 
The transformation of the baryon octet and pentaquark antidecuplet 
${\cal P}$ including $\Theta^+$ ($I=0$) can be chosen as 
\[
{B^i}_j \rightarrow {U^i}_a\, {B^a}_b \,{U^{\dagger\,b}}_j,\quad
\mathcal{P}_{ijk} \rightarrow P_{abc}\, 
{U^{\dagger\,a}}_i \,
{U^{\dagger\,b}}_j \,
{U^{\dagger\,c}}_k, 
\]
where all the indices are for $\textrm{SU}(3)$ flavor.
The pentaquark baryons are related to ${\cal P}_{abc} = {\cal P}_{(abc)}$ by,
for example,
${\cal P}_{333}=\Theta^+$,
${\cal P}_{133}=\frac{1}{\sqrt{3}}\widetilde{N}^0$,
${\cal P}_{113} = \frac{1}{\sqrt{3}}\widetilde{\Sigma}^-$, and
${\cal P}_{112} = \frac{1}{\sqrt{3}}\Xi_{3/2}^{-}$.
Then, one can define a covariant derivative ${\cal D}_\mu$,
which transforms as ${\cal D}_\mu B \rightarrow U {\cal D}_\mu B U^{\dagger}$, 
by
\[
{\cal D}_\mu B = \partial_\mu B + [ V_\mu , B ] .
\]

Chiral symmetry is explicitly broken by non-vanishing current-quark masses
and electromagnetic interactions. The former can be included by regarding 
the quark-mass matrix $m = \textrm{diag}\,( m_u , m_d , m_s )$ as a spurion 
with transformation property 
$m \rightarrow L m R^{\dagger} = R m L^{\dagger}$. 
It is more convenient to use $\xi m \xi + \xi^{\dagger} m \xi^{\dagger}$,
which transforms as an $\textrm{SU}(3)$ octet. 
Electromagnetic interactions can be included by introducing photon field 
${\cal A}_\mu$ and its field strength tensor
$F_{\mu \nu} = \partial_\mu {\cal A}_\nu - 
\partial_\nu {\cal A}_\mu$:
\begin{subequations}
\begin{eqnarray}
{\partial}_\mu \Sigma & \rightarrow & 
{\cal D}_\mu \Sigma
\equiv
\partial_\mu \Sigma + i e {\cal A}_\mu [ Q , \Sigma ],
\\
V_\mu & \rightarrow & V_\mu + \mbox{$\frac{ie}{2}$} {\cal A}_\mu
( \xi^{\dagger} Q \xi + \xi Q \xi^{\dagger} ), 
\\
A_\mu & \rightarrow & A_\mu - \mbox{$\frac{e}{2}$} {\cal A}_\mu
( \xi^{\dagger} Q \xi - \xi Q \xi^{\dagger} ), 
\end{eqnarray}
\end{subequations}
where $Q \equiv {\rm diag}~(2/3, -1/3, -1/3)$ is the electric-charge 
matrix for light quarks ($q=u,d,s$). 

Now it is straightforward to construct a chiral lagrangian with lowest order
in derivative expansion. 
The parity and charge-conjugation symmetric chiral lagrangian is given by  
\begin{equation}
\mathcal{L} = \mathcal{L}_{\Sigma} + \mathcal{L}_{B} + \mathcal{L}_{\cal P},
\end{equation}
where 
\begin{subequations}
\begin{eqnarray}
{\cal L}_{\Sigma} & = & {f_\pi^2 \over 4}\,\textrm{Tr} \big[ 
{\cal D}_\mu \Sigma^{\dagger} {\cal D}^{\mu} \Sigma
-2\mu m ( \Sigma + \Sigma^{\dagger} )
\big], 
\\
{\cal L}_{B} & = & 
\textrm{Tr}\,\overline{B} (i \mathcal{D}\!\!\!\!/-m_B) B 
+
D\,\textrm{Tr}\,\overline{B} \gamma_5 \{A\!\!\!/,B\}
\nonumber\\
&&
+
F\,\textrm{Tr}\,\overline{B} \gamma_5 [A\!\!\!/,B],
\\
{\cal L}_{\cal P} & = & 
\overline{\mathcal{P}}(i \mathcal{D}\!\!\!\!/-m_\mathcal{P}) \mathcal{P}
+\mathcal{C}_{\mathcal{P}N}
\big(\overline{\mathcal{P}}\Gamma_P A\!\!\!/ B
    +\overline{B}          \Gamma_P A\!\!\!/ \mathcal{P}
\big)
\nonumber\\
&&
+\mathcal{H}_{\mathcal{P}N}
\overline{\mathcal{P}}\gamma_5 A\!\!\!/ \mathcal{P},
\end{eqnarray} 
\end{subequations}
where $P$ is the parity of $\Theta^+$, 
$\Gamma_{+} = \gamma_5$, and $\Gamma_{-} = 1$, and %, respectively. 
$m_{\cal P}$ is the average of the pentaquark 
decuplet mass.

The Gell-Mann-Okubo formulae for pentaquark baryons will be obtained from
\begin{equation}
{\cal L}_{m} = \alpha_m \,\overline{\cal P} ( \xi m \xi + \xi^{\dagger} m 
\xi^{\dagger} ) {\cal P}. 
\end{equation}
Expanding this, we get the mass splittings 
$\Delta m_i\equiv m_i-m_{\mathcal{P}}$ within the antidecuplet: 
\begin{subequations}
\begin{eqnarray}
\Delta m_\Theta & = & 2 \alpha_m m_s,
\\
\Delta m_{\widetilde{N}} & = & 
\alpha_m \left( 2 \hat{m} + 4 m_s \right)/3, 
\\
\Delta m_{\widetilde{\Sigma}} & = &  \alpha_m \left( 4 \hat{m} + 
2 m_s \right)/3, 
\\
\Delta m_{\Xi_{3/2}} & = &  2~ \alpha_m \hat{m}, 
\end{eqnarray}
\end{subequations}
where $\hat{m} = m_u = m_d$ ignoring small isospin-breaking effects.
If the newly observed state at a mass $1862 \pm 2$ MeV is identified 
as $\Xi_{3/2}$, we find
\begin{equation}
m_{\widetilde{N}}  =  1647 \,{\rm MeV},\quad
m_{\widetilde{\Sigma}}  =  1755\,{\rm MeV}.
\end{equation}

Recently Jaffe and Wilczek suggested there could be an ideal mixing between 
pentaquark antidecuplet $\mathcal{P}_{abc}$  and pentaquark octet 
${\mathcal{O}^a}_b$ with the 
same parities \cite{Jaffe:2003sg}. 
This idea has been generalized further by other groups 
\cite{Diakonov:2003jj,Oh:2003fs}.
In chiral lagrangian approach, such a general mixing arises from 
\begin{equation}
\beta_m\left[ 
\overline{\cal P} ( \xi m \xi + \xi^{\dagger} m \xi^{\dagger} ) {\cal O}
+ \overline{\cal O} ( \xi m \xi + \xi^{\dagger} m \xi^{\dagger} ) {\cal P} 
\right].
\end{equation}
Expanding this leads to 
\begin{equation}
{\cal L} = 
\mathcal{B}_m\big[
\overline{p} \widetilde{N}^+ - \overline{n} \widetilde{N}^0
+ \overline{\Sigma^0} \widetilde{\Sigma}^0 
- \overline{\Sigma^-} \widetilde{\Sigma}^- 
+ \overline{\Sigma^+} \widetilde{\Sigma}^+ + \,\textrm{H.c.} \big],
\label{eq:Lm}  
\end{equation}
where 
$\mathcal{B}_m=2 \beta_m ( m_s - \hat{m} )/\sqrt{3}$
and
we borrowed baryon-octet notation for pentaquark octet 
states in Eq.~(\ref{eq:Lm}).
Note that the relative sign between $\overline{n} \widetilde{N}^0$ and 
$\overline{p} \widetilde{N}^+$ 
(and also $\overline{\Sigma^0} \widetilde{\Sigma}^0$) is negative, unlike 
the case in Ref.~\cite{Diakonov:2003jj}. This is due to the 
$\overline{\bm{10}}$ nature of the $\mathcal{P}$.  
One could write down the same mixing between pentaquark antidecuplet 
${\cal P}_{abc}$ and the ordinary baryon octet $B$, but such terms will be  
highly suppressed compared to the above term, since it is a mixing between 
$qqq$ and $qqqq\bar{q}$.

Finally the baryon decuplet can only couple to pentaquark octet 
$\mathcal{O}$, but not to pentaquark antidecuplet ${\cal P}$,  since 
$\bm{10 \otimes 8 \otimes 10}$ does not contain $\textrm{SU}(3)$ singlet. 
This implies that $N(1440)$ or $N(1710)$ cannot be pure pentaquark 
antidecuplets, because they have substantial branching ratios into 
$\Delta \pi$ final states. They could be mixed states of pentaquark octet
and pentaquark antidecuplet, and their productions and decays will be more 
complicated than pure antidecuplet case. Since the current data on baryon 
sectors are not enough to study such mixings in details, we do not pursue 
the mixing further in the following. 

Parameters in the above lagrangian are taken to have the following numerical
values: 
$m_B \approx 940\,$MeV is the nucleon mass, 
 $D \approx -0.81$ and $F \approx -0.47$ at tree level, and 
we assume $\hat{m} = 0$ and $m_\eta^2 = (4/3)~m_K^2$. 

The coupling ${\cal C}_{\mathcal{P}N}$ is determined
from the decay width $\Gamma_{\Theta}$ of the $\Theta^+$,  
which is dominated by $K^+ n$ and $K^0 p$ modes as
$\Gamma_{\Theta}/2=\Gamma_{\Theta^+\to K^+ n}%
=\Gamma_{\Theta^+\to K^0 p}$:
\begin{equation}
\Gamma_{\Theta}=
\frac{\mathcal{C}^2_{\mathcal{P}N}|\textbf{p}^*|}{8\pi f^2 m_\Theta^2}
(m_\Theta\pm m_B)^2~[(m_\Theta\mp m_B)^2-m_K^2],
\end{equation}
where $\textbf{p}^*$ is the kaon momentum in the $\Theta^+$
rest frame and the signs are for $P(\Theta^+)=\pm 1$, respectively.
Then the $\mathcal{C}_{\mathcal{P}N}$ is determined as
\[
\mathcal{C}^2_{\mathcal{P}N}(P=+,-)
=(2.7,0.90)\times \Gamma_{\Theta}/\textrm{GeV}.
\]
Cahn and Trilling \cite{Cahn:2003wq} argues that 
$\Gamma_{\Theta}= (0.9 \pm 0.3)$MeV using the DIANA results 
\cite{Barmin:2003vv}. 
We present our results proportional to 
the $\mathcal{C}_{\mathcal{P}N}^2$
rescaled by the factor $1\,\textrm{MeV}/\Gamma_{\Theta}$. 
Such a small $\mathcal{C}_{\mathcal{P}N}$ can be understood as following:
this coupling is related to the matrix element of hadronic axial vector 
current operator (with zero baryon number) between a pentaquark baryon and 
an ordinary baryon. Since two states have different number of valence quarks,
this matrix element should be highly suppressed compared to the ordinary
axial vector coupling $(D+F)$ or ${\cal H}_{\mathcal{P}N}$. 

The coupling ${\cal H}_{\mathcal{P}N}$ is also unknown, and determines 
transition rates between pentaquark antidecuplets with pion or kaon emission.
Unfortunately, such decays are all kinematically forbidden, 
and cannot be used to fix ${\cal H}_{\mathcal{P}N}$. 
However, we expect that ${\cal H}_{\mathcal{P}N} = O(1)$, without 
any suppression as in $\mathcal{C}_{\mathcal{P}N}$.
With this remark in mind, we will assume ${\cal H}_{\mathcal{P}N}$ can vary 
between $-4$ and 4 in the numerical analysis to be as general as possible.

\begin{figure}
\includegraphics[height=2.0cm]{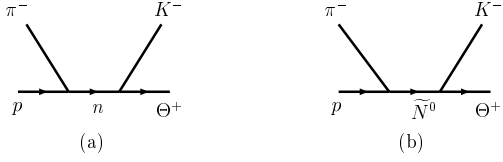}%
\caption{\label{fig1} Feynman diagrams for $\pi^- p \rightarrow K^- \Theta^+$.}
\end{figure}

3. \textit{$\pi^- p \rightarrow K^- \Theta^+$} ---
Let us first consider $\pi^- p \rightarrow  K^- \Theta^+$ as applications of
our chiral lagrangian for pentaquark baryons. In Figs.~\ref{fig1} (a) and 
(b), we show the relevant Feynman diagrams.  Note that only Fig.~\ref{fig1} 
(a) was considered in the literature. However, there is an $s-$channel 
$\widetilde{N}^0 (1647)$ exchange diagram [ Fig.~\ref{fig1} (b) ] 
in our chiral lagrangian, since $\Theta^+$ is not an $\textrm{SU}(3)$ 
singlet, but belongs to the antidecuplet. 
Therefore one has to keep both Figs.~\ref{fig1} (a) and (b) in order 
to get an amplitude with correct $\textrm{SU}(3)$ flavor symmetry. 
 
\begin{figure}
\begin{tabular}{cc}
{\includegraphics[height=4cm]{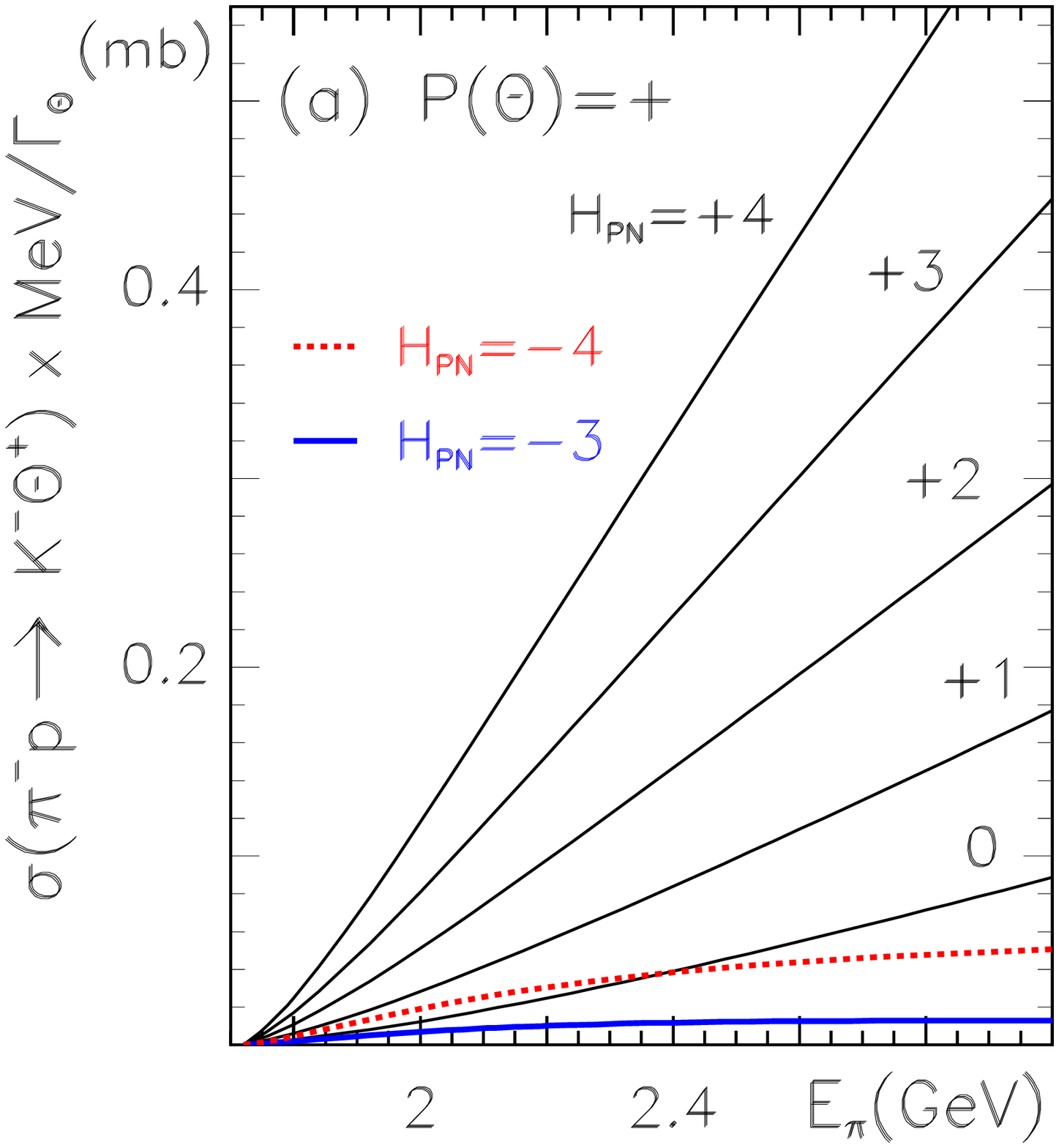}}&
{\includegraphics[height=4cm]{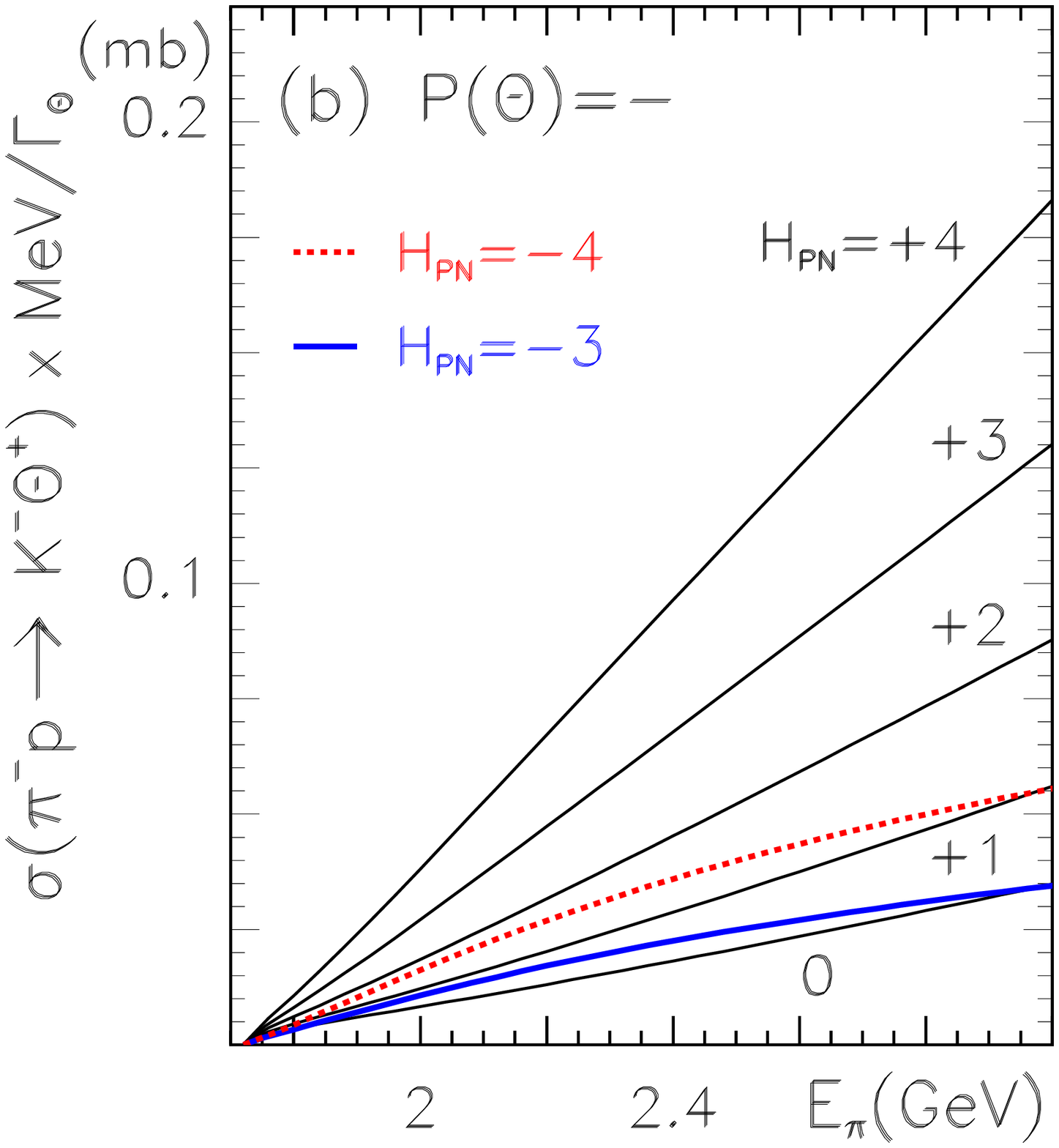}}
\end{tabular}
\caption{\label{fig2} 
Cross sections for 
$\pi^- p \rightarrow K^- \Theta^+$ 
for selective values of $\mathcal{H}_{\mathcal{P}N}$ between $-4$ and $+4$,  
in the proton rest frame: (a) $P ( \Theta ) = +1$ and 
(b) $P ( \Theta ) = +1$.
}
\end{figure}

The amplitude for $\pi^- p \rightarrow K^- \Theta^+$ is given  by
\begin{eqnarray}
\mathcal{M} &=&
\frac{\mathcal{C}_{\mathcal{P}N}}{2f^2}
\bar{u}_{\Theta^+}
\bigg(
\Gamma_P / \! \! \! p_{K^-}
 \frac{D+F}{  /\! \! \! p_{\pi^-} + /\! \! \! p_{p} -m_B}
\gamma_5/ \! \! \! p_{\pi^-}
\nonumber\\
&&\quad
-
\gamma_5 / \! \! \! p_{K^-}
\frac{{\cal H}_{\mathcal{P} N}/3}
               { / \! \! \! p_{\pi^-} + /\! \! \! p_{p} - m_{\widetilde{N}}}
\Gamma_P/ \! \! \! p_{\pi^-}
\bigg)
    u_p.  
\end{eqnarray}

We show the total cross section, 
rescaled by the factor $\Gamma_{\Theta}/\textrm{MeV}$,
as functions of pion energy $E_\pi$ at the proton rest frame in 
Fig.~\ref{fig2} depending on the $\Theta^+$ parity,
and varied $-4 \leq \mathcal{H}_{\mathcal{P} N} \leq 4$.
Note that the sign of ${\cal H}_{\mathcal{P} N}$ is very important.
If ${\cal H}_{\mathcal{P} N}>0\,(<0)$, two contributions will have 
constructive (destructive) interference. Thus our results differ  
from the previous results where only the $n$ contribution was included.
Also the cross section is sensitive to ${\cal H}_{\mathcal{P}N}$, and
may be useful to fix ${\cal H}_{\mathcal{P}N}$. 
Following the spirit of chiral perturbation theory at lowest order,
we did not include model-dependent form factors, keeping
in mind that our results get unreliable as the $E_\pi$ gets 
as large as $\sim 2.5$ GeV.

Note that the even parity and the odd parity cases can be distinguished 
from the cross section for the parity-odd case is smaller than that for 
the parity-even case. 

4. \textit{Photo-production of $\Theta^+$} ---
In order to study photoproduction of $\Theta^+$ on nucleons, we need to 
know the magnetic dipole interaction terms. For the nucleon octet,
\[
{\cal L} = -{e \over 4 m_B}\textrm{Tr}\bigg[
\overline{B} \sigma_{\mu\nu} F^{\mu\nu} 
\big(\kappa_D\{ Q , B \} +\kappa_F [ Q , B ]\big)
 \bigg].
\]
The anomalous magnetic moments of nucleons are 
\[
\kappa_p = \kappa_F + \mbox{$\frac{1}{3}$}\, \kappa_D,\quad
\kappa_n = -\mbox{$\frac{2}{3}$}\, \kappa_D,
\]
at tree-level chiral lagrangian. Using 
$\kappa_p =  1.79$ and 
$\kappa_n = -1.91$, 
we get 
$\kappa_D =  2.87$ and 
$\kappa_F =  0.836$. 
For the pentaquark baryon ${\cal P}$, the relevant term is 
\[
- \frac{e\kappa_{\cal P}}{4 m_{\mathcal{P}}}\,q_i \overline{\cal P}_{i} 
\sigma_{\mu \nu}~F^{\mu \nu}~{\cal P}_{i} \rightarrow 
-\frac{e\kappa_{\Theta}}{4 m_{\Theta}}\,
\overline{\Theta^+} \sigma_{\mu\nu} F^{\mu\nu} \Theta^+. 
\]
We expect that 
$| \kappa_\Theta(\equiv\kappa_{\cal P})|\approx|\kappa_D|\approx|\kappa_F|$.
On the other hand, a calculation in soliton picture predicts that 
$\kappa_{\Theta} \approx 0.3$, which is rather small \cite{Kim:2003ay}. 
We vary $\kappa_{\Theta}$ between $-1$ and 1.
We ignore  transition magnetic moments between nucleon octet
and pentaquark antidecuplet, since this transition involves $qqq$ and 
$qqqq\bar{q}$. 

\begin{figure}
\includegraphics[height=2cm]{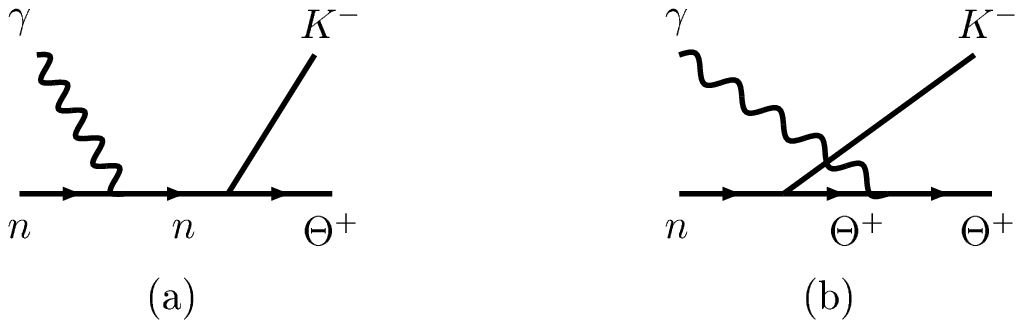}
\includegraphics[height=2cm]{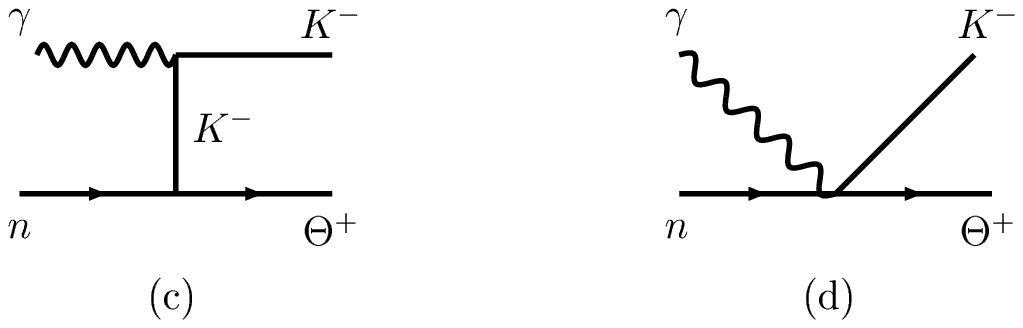}
\caption{\label{fig3} 
Feynman diagrams for $\gamma n \rightarrow K^- \Theta^+$.}
\end{figure}

\begin{figure}
\begin{tabular}{cc}
{\includegraphics[height=4cm]{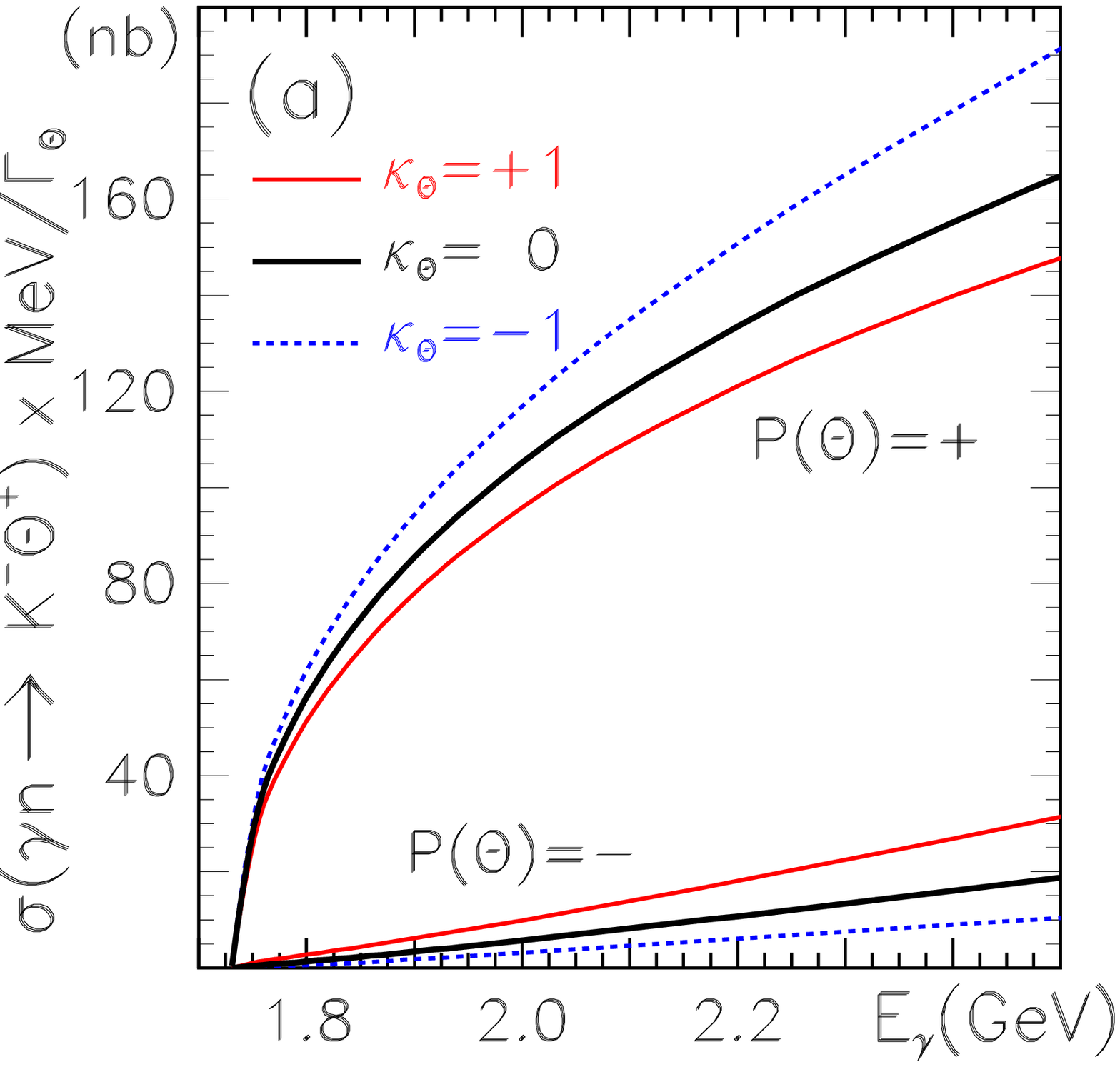}}&
{\includegraphics[height=4cm]{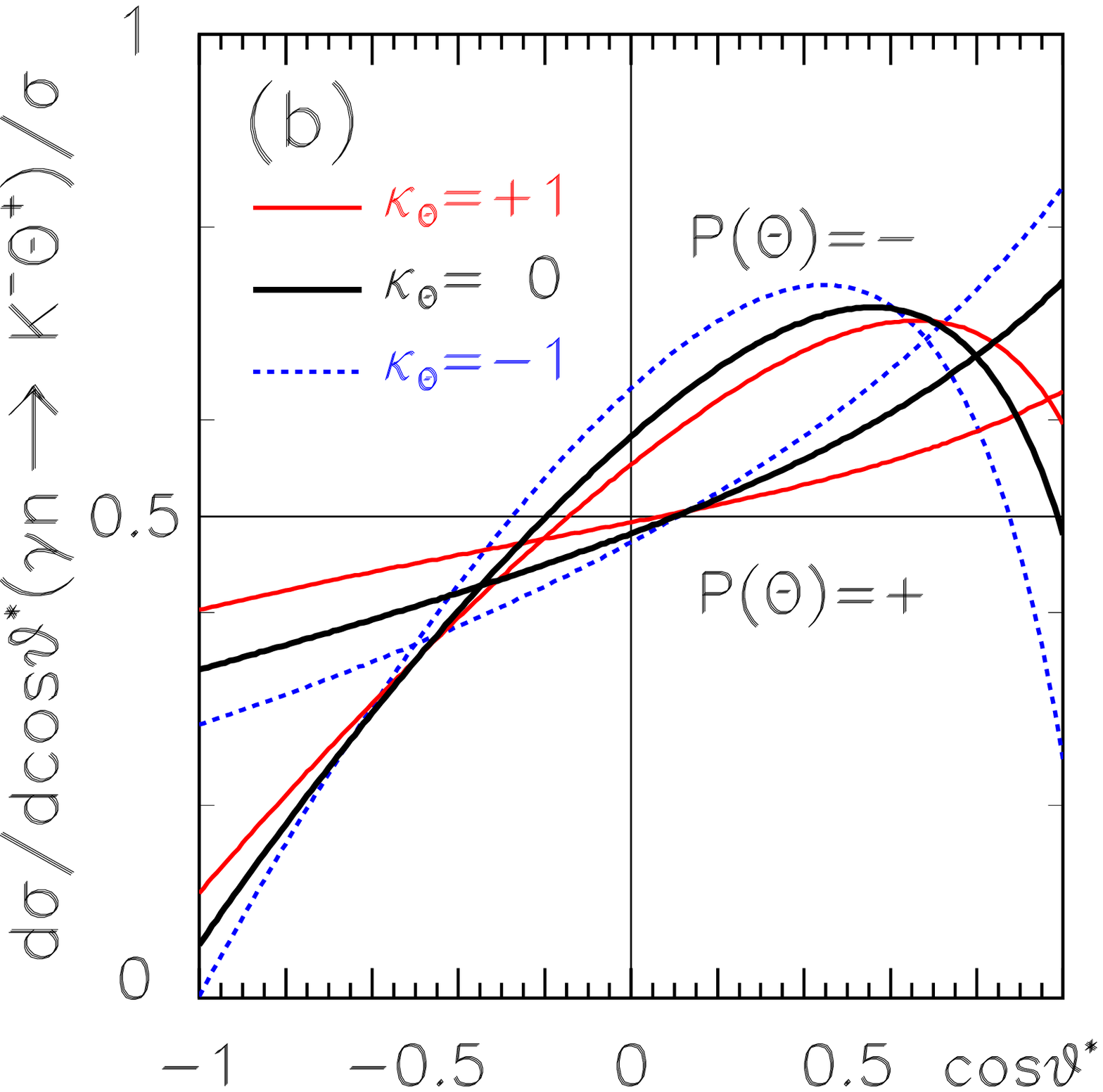}}
\end{tabular}
\caption{\label{fig4} (a) Cross sections for 
$\gamma n \rightarrow K^- \Theta^+$ and (b) the angular distribution for 
$E_{\gamma} = 2$ GeV in the center of momentum  frame.
}
\end{figure}

The relevant Feynman diagrams for $\gamma n \rightarrow K^- \Theta^+$ 
are shown in Fig.~\ref{fig3}. One salient feature of our approach based on 
chiral perturbation theory is the existence of a contact term for 
$\gamma K^- n \Theta^+$ vertex [Fig.~3 (d)] that arises 
from the $\mathcal{C}_{\mathcal{P}N}$ term in Eq.~(4c) with Eq.~(2c),
which is necessary to recover $\textrm{U}(1)_{\textrm{em}}$ gauge 
invariance within spontaneously broken global chiral symmetries. 
The resulting amplitude is 
\begin{subequations}
\begin{equation}
\mathcal{M}  =  \frac{e \mathcal{C}_{\mathcal{P}N}}{\sqrt{2} f}\,
\epsilon_\mu \,\bar{u}_{\Theta^+} F^\mu u_n, 
\end{equation}
\begin{eqnarray}
F^\mu & = & 
-\Gamma_P    {p\!\!\!/}_{K}
\frac{\kappa_n}{4 m_B}\,
\frac{1}{  {p\!\!\!/}_{\gamma} + {p\!\!\!/}_n  - m_B } 
    [ \gamma^\mu , {p\!\!\!/}_{\gamma} ] 
+ \Gamma_P \gamma^\mu
\nonumber  \\
& + & \bigg( \gamma^\mu - \frac{\kappa_\Theta}{4 m_\Theta}\,
             [ \gamma^\mu , {p\!\!\!/}_{\gamma} ] 
      \bigg)\,
     \frac{1}{ {p\!\!\!/}_n -{p\!\!\!/}_K - m_\Theta }\,
     \Gamma_P  {p\!\!\!/}_K
\nonumber  \\ 
& - & \frac{( 2 p_K - p_\gamma )^\mu}{ ( p_K - p_\gamma )^2 - m_K^2}\,
\Gamma_P   ({p\!\!\!/}_K -{p\!\!\!/}_{\gamma}).
\end{eqnarray} 
\end{subequations}

The cross sections and the angular distributions in the center of momentum
frame are shown in Fig.~\ref{fig4} (a) and (b). Note that the
parity-even case has larger cross section, and has a sharp rise near the 
threshold. The angular distribution shows that the forward/backward 
scattering is suppressed in the negative parity case, whereas the forward 
peak is present in the positive parity case. Therefore the angular 
distribution could be another useful tool to determine the parity of 
$\Theta^+$. 
Therefore, once $\mathcal{C}_{\mathcal{P}N}^2$ is determined from 
$\Gamma_\Theta$, one could determine the parity of $\Theta^+$, and make 
a rough estimate of $\kappa_\Theta$ from the photoproduction cross section. 

5. \textit{Including light vector mesons} ---
One can also introduce light vector mesons $\rho_{\mu}$, which transforms as
\begin{equation}
\rho_\mu (x) \rightarrow U (x) \rho_\mu (x) U^{\dagger} (x) + 
U (x) \partial_\mu U^{\dagger} (x),
\end{equation}
under global chiral transformations \cite{Bando:1985rf}. 
Then $\rho_\mu (x)$ transforms as a gauge field under local 
$\textrm{SU}(3)$'s defined by Eq.~(1), as $V_\mu$ does. 
The covariant derivative $\mathcal{D}_\mu$ can be defined using $\rho_\mu$
instead of $V_\mu$. 
Note that $( \rho_{\mu} - V_\mu )$ has a simple transformation property  
under chiral transformation: 
\[
 (\rho_{\mu} - V_\mu ) \rightarrow U(x)  
 (\rho_{\mu} - V_\mu )
U^{\dagger} (x), 
\]
and it is straightforward to construct chiral invariant lagrangian using 
this new field. In terms of a field strength tensor $\rho_{\mu\nu}$,
%For example, 
\begin{eqnarray*}
{\cal L}_{\rho} & = & -\frac{1}{2}~ {\rm Tr} ( \rho_{\mu\nu} \rho^{\mu\nu} ) 
+ \frac{1}{2}~m_{\rho}^2 ~ {\rm Tr} ( \rho_{\mu} - V_\mu )^2 
\\
& + & \alpha \left[ \overline{\mathcal{P}} 
         ( \rho \!\!\!/ - V \!\!\!\!\!/ \, ) B 
 + \overline{B} 
    ( \rho \!\!\!/ - V \!\!\!\!\!/ \, ) {\mathcal{P}}  \right] 
+ ...
\end{eqnarray*}
It is important to notice that $N \Theta^+ K^*$ coupling should be highly
suppressed, since it can appear only in combination of 
$( \rho_{\mu} -  V_\mu )$, which vanishes in the low-energy limit.
In other words, the low-energy theorem is violated if one includes only 
$n \Theta^+ K^*$ diagram, without including the $n \Theta^+ K \pi$ contact 
term arising from the $ V \!\!\!\!\!/$ term. Therefore, one should be cautious 
about claiming that the $K^*$ exchange is important in 
$\pi^- p \rightarrow K^- \Theta^+$. 
Detailed numerical analysis of vector-meson exchange is straightforward, 
but beyond the scope of the present work and will be pursued elsewhere 
\cite{work}. 

6. \textit{Conclusion} ---
In conclusion, we constructed a chiral lagrangian involving pentaquark 
baryon antidecuplet and octet, the ordinary nucleon octet and Goldstone 
bosons. 
%Parameters in our chiral lagrangian are important to describe the 
%masses, mixings and dynamics of pentaquark baryons. Lacking a reliable 
%calculation tool from underlying QCD, we fix thm from experimental data. 
Using this lagrangian, we derived the Gell-Mann-Okubo formula and the mixing
between the pentaquark antidecuplet and pentaquark octet. We also 
calculated the cross sections for 
$\pi^- p \rightarrow K^- \Theta^+$ and $\gamma n \rightarrow K^- \Theta^+$ 
for $J^P=\frac{1}{2}{}^{\pm}$. In particular, we emphasized that it is very 
important to respect chiral symmetry properly in order to get correct
amplitudes for these processes. All these observables depend on parameters
in our chiral lagrangian, which have relations with underlying QCD, but 
are uncalculable from QCD at present. 
Photo-production data will be particularly useful in identifying the 
parity of $\Theta^+$, because the threshold behavior of the cross section 
and its angular distributions strongly depend on the parity.
Once the coupling $\mathcal{C}_{\mathcal{P}N}$ is 
determined from the decay width of $\Theta^+$, then the parity and other
couplings $\mathcal{H}_{\mathcal{P}N}$ and $\kappa_\Theta$ could be determined
from the hadro/photo-production cross sections for $\Theta^+$.
It is straightforward to apply our approach to other
related processes such as $\gamma p \rightarrow K^0 \Theta^+$ or $K^+ p 
\rightarrow \pi^+ \Theta^+$, or other pentaquark baryons.  
Finally we have outlined how to  incorporate the vector meson degrees of 
freedom in our scheme, the details of which will be discussed in the 
separate publication \cite{work}.

\begin{acknowledgments}
This work is supported in part by KOSEF through CHEP at Kyungpook National
University and by the BK21 program.
The research of JL~in the High Energy Physics Division
at Argonne National Laboratory is supported by
the U.~S.~Department of Energy, Division of High Energy Physics, under
Contract W-31-109-ENG-38. 
JL thanks KAIST and KIAS for their hospitality during this work. 
\end{acknowledgments}


\end{document}